\documentclass[preprint,prd,nofootinbib,tightenlines,amsmath]{revtex4}
\usepackage{bbm}
\usepackage{amsfonts}
\usepackage{mathrsfs}
\usepackage{multirow,array}
\usepackage{graphicx}
\usepackage{bm}

\begin{document}
\baselineskip=15pt \parskip=5pt

\vspace*{3em}

\title{Search for pseudoscalar-mediated WIMPs in $t \rightarrow c$ transitions with missing energy}

\author{Lian-Bao Jia}
\email{jialb@mail.nankai.edu.cn}

\affiliation{School of Science, Southwest University of Science and Technology, Mianyang
621010, P. R. China \\}

\begin{abstract}

The recent astronomical observation of GeV gamma-ray excess from the Galactic Center was suggested due to a $b \bar{b}$ mode in tens GeV WIMP (weakly interacting massive particle) pair annihilations, and this mode was also explored by the new dwarf galaxy observation. Considering the case the WIMP pair mass below top quark mass, a pseudoscalar $\phi$ is studied in this article, which mediates the interactions between the standard model fermions and fermionic WIMPs, and neutral flavor-changing interactions in standard model fermion sectors. The $b \bar{b}$ mode is favored in WIMP pair annihilations, while the WIMP-nucleus scattering is highly suppressed in direct detection. Alternative schemes of $t \rightarrow c$ decay and single top production are employed to search WIMPs. Assuming the mass of WIMP around 5-60 GeV and with the reasonable inputs by the constraints, the branching ratio $\mathcal {B}_{t\rightarrow c \bar{\chi} \chi}$ of a top quark decaying into a charm quark and a WIMP pair is derived of order $10^{-8} - 10^{-5}$, thus careful studies in the future on top-physics may help to gain a better understanding of WIMPs.

\end{abstract}

\maketitle

\section{Introduction}

It is believed that the major constituent of the matter in the universe is dark matter (DM),
and one of the most favored DM candidates are WIMPs (weakly interacting massive particles).
The recent results from LUX \cite{Akerib:2013tjd} and SuperCDMS \cite{Agnese:2013jaa,Agnese:2014aze} set stringent constraints on WIMPs in direct search, and meanwhile the collider physics is providing complementary conditions to narrow possible survival ranges for WIMP models.
We definitely know that the WIMP-nucleus scattering cross section may be suppressed by $q^2$, $v^2$, or $J$,  where $q$ is the momentum transfer, $v$ is the WIMP relative velocity, and $J$ is the total spin of the target nucleus, thus so far no firm signal has been observed in DM direct detection. We expect that
the WIMP signals will be detected above the neutrino irreducible background in the next decades by the ultimate DM direct detectors (see e.g.~\cite{Freytsis:2010ne,Baudis:2012ig,Billard:2013qya,Ruppin:2014bra,Baudis:2014naa,Blum:2014dca} for more). As a matter of fact, even though the signals are below the neutrino background, with today's sophisticated facilities and understanding of neutrinos, we will still be possible to distinguish them from the background.

The highly suppressed WIMP-nucleus interaction may make the WIMPs to evade our direct detection, so some alternative ways to detect WIMPs are needed. An important means is obviously the accelerator physics. If at the high energy accelerators, such as LHC, the WIMPs which are definitely not the standard fermions or bosons are produced as missing energy,  we could identify them and declare a success. But since the background of LHC is too complicated, it would be hard to dig out missing energy from the messy products and re-establish the concerned events. An alternative detection was proposed the authors of Ref. \cite{Bird:2004ts} suggested that  rare decays of B meson might be feasible to search WIMPs of a few GeV and sub-GeV in $b\rightarrow s$ transitions with missing energy. If the masses of WIMPs are to heavy to be detected in B-decays, a natural extension of the idea is to the top quark decays \cite{He:2007tt,Li:2011ja}.

The gamma-ray emissions from DM dense regions may provide the information about WIMPs, e.g. the Galactic Center gamma rays \cite{Goodenough:2009gk,Hooper:2010mq,Abazajian:2010zy,Boyarsky:2010dr,Hooper:2011ti,Abazajian:2012pn,Gordon:2013vta}. Recent studies of the excess of $\sim$1-3 GeV gamma rays \cite{Abazajian:2014fta,Daylan:2014rsa,Calore:2014xka,Agrawal:2014oha,Calore:2014nla,Calore:2015nua} and antiproton spectrum \cite{Hooper:2014ysa} from the region surrounding the Galactic Center indicate that the excess can be interpreted as annihilating WIMPs, for WIMPs of $35\sim 51$ GeV mainly annihilating to $b\bar b$ pairs with a cross section of $\sim$ $10^{-26}$ cm$^3/$s $(b\bar{b})$.
These results are at the order of the expectation values of the thermal freeze-out WIMPs \cite{Steigman:2012nb} and are consistent with the prediction of the model where a pseudoscalar boson is the interaction messenger \cite{Boehm:2014hva,Ipek:2014gua,Ghorbani:2014qpa,Dolan:2014ska,Berlin:2015sia,Kozaczuk:2015bea,Berlin:2015wwa} (for general discussions, see e.g. Refs. \cite{Berlin:2014tja,Izaguirre:2014vva}). Moreover, as pointed in Ref. \cite{Hooper:2015ula}, the interpretation of the gamma ray excess can be tolerated by the limits from the new dwarf galaxy \cite{Ackermann:2015zua} given by the Fermi-LAT Collaboration. That is an encouraging piece of information and may hint a direction to search for WIMPs, if the allegation about WIMP annihilation causing the excess is valid. In this work, the limits from the new dwarf galaxy will be taken into consideration.

An optimistically alternative scheme to detect the WIMPs is via the flavor-changing processes of $t \rightarrow c$ decay and single top production (see e.g. Refs. \cite{He:2007tt,Li:2011ja,Andrea:2011ws,Kamenik:2011nb,Blanke:2013zxo} about the missing energy in $t-c,u$ transitions). When a WIMP pair with mass below the top quark mass (e.g. tens GeV), the $t \bar{c}$ ($\bar{t} c$) mode in WIMP pair annihilations will be negligible, and the main SM products are from other modes, e.g. the $b \bar{b}$ mode as indicated by the observation of anomalous gamma ray excess. This is the thing of concern and study in this paper. In this case, the channel of $t \rightarrow c$ decay may be an ideal to search WIMPs.

Now, the crucial point is to reasonably estimate the rate of the rare decay because it is the key whether the channel can be observed at LHC and the future facilities, such as the planned
ILC \cite{Baer:2013cma} or other proposed top-factory. Indeed, the estimate of the rate depends on the models of WIMPs, namely, for different models, the estimated rate might be quite apart.
It is also well known that DM cannot interact with the regular particles via Standard Model (SM), but something beyond SM. What is that, so far nobody
is sure, however, there are many plausible models available, and each of them undergoes stringent test of astronomical observation and accelerator experiments, as the corresponding
parameter space is rigorously constrained.

In this work, we carry out our calculation based on the WIMP model where a new pseudoscalar mediates the interaction between DM and SM, and neutral flavor-changing interactions are introduced for SM fermions.

This work is organized as follows. After this introduction, we introduce the Lagrangian for the interactions where a new pseudoscalar boson is responsible
for the annihilation of WIMPs and production of the SM partlces (say, $b\bar b$), and then formulate its contributions to $t - c$ transitions with the missing energy. We also evaluate the annihilation cross section of WIMPs. Next, the numerical results are presented along with all the input parameters. The last section is devoted to our conclusion and a brief discussion.

\section{The missing energy in $t - c$ transitions}

As discussed in the introduction, in this work, we employ the model where a flavor-changing pseudoscalar bridges between the SM fermions and fermionic WIMPs.
In this section, we will consider a few neutral flavor-changing processes where the top quark and the new pseudosclar boson  are involved.

\subsection{The interaction between SM and WIMPs}

Here we present the concerned interaction, by which the SM FCNC (flavor changing neutral current) processes are realized via exchanging a neutral pseudoscalar boson ${\phi}$ at tree level. The effective couplings of ${\phi}$ to quarks and leptons are depicted as
\begin{eqnarray}
\mathcal {L}_{SM}^{\,i} = -\lambda_{\,{q}'q}\bar{q}' i \gamma_5 q \phi- \lambda_{\,{l}'l}^{}\bar{l}' i \gamma_5 l {\phi}+h.c.\,,
\end{eqnarray}
where $\lambda_{\,{q}'q}^{}$, $\lambda_{\,{l}'l}^{}$ are dimensionless parameters of ${\phi}$ corresponding to quarks and leptons respectively. In Cheng and Sher ansatz~\cite{Cheng:1987rs}, the flavor-changing interactions are correlated with the geometric mean values of the two relevant fermion masses. Following Ref. \cite{Cheng:1987rs}, we set the flavor-changing parameters in similar forms, that is,
\begin{eqnarray}
\lambda_{\,{q}'q}^{}\sim \theta \frac{\sqrt{2 m_{q'}m_q}}{\upsilon}\,, \,\,\,\, \lambda_{\,{l}'l}^{} \sim \theta \frac{\sqrt{2 m_{l'}m_l}}{\upsilon}\,, \label{c-SM}
\end{eqnarray}
where $\upsilon / \sqrt{2}$ is the electroweak vacuum expectation value, as $\upsilon \approx 246$ GeV, and $\theta$ is a  dimensionless parameter which is much smaller than unity. A solution to the mass and mixing hierarchies in quark and lepton sectors was explored in Ref. \cite{Froggatt:1978nt} in Froggatt-Nielsen mechanism (see e.g. Ref. \cite{Dery:2014kxa} a recent discussion about flavor violations).

The vertex for the fermionic WIMP-pseudoscalar is $-i\lambda_D \gamma_5$, and $\lambda_D$ is the effective coupling constant, in the perturbative limit $\lambda_D < \sqrt{4 \pi }$. Thus, the total Yukawa couplings between a pseudoscalar field $\phi$ and SM fermions, fermionic WIMPs are
\begin{eqnarray}
\mathcal {L}_{SM+DM}^{\,i} = -i \lambda_{\,{q}'q}^{} \bar{q}' \gamma_5 q {\phi}-i \lambda_{\,{l}'l}^{} \bar{l}' \gamma_5 l {\phi}-i\lambda_D \bar{\chi} \gamma_5 \chi {\phi} +h.c.\,, \label{SM-DM}
\end{eqnarray}
where $\chi$ is the WIMP field.

The scattering operator between WIMPs and target nucleus is of the form $\bar{\chi} \gamma_5 \chi \bar{q}' \gamma_5 q$. This interaction structure is spin-dependent, and the scattering cross section is suppressed by $q^4$, as discussed in~\cite{Freytsis:2010ne,Berlin:2014tja}. Thus, for WIMPs interacting via this type
Lagrangian, the WIMPs would evade the direct detection at the present
underground experiment. Therefore, if it is the case, the possible way to
identify them is evaluating the contributions of $\phi$ to the processes of $t \rightarrow c$ decay, or single top production.

\subsection{The flavor-changing neutral transitions of $t $}

By Eq.~(\ref{c-SM}), it is found that the coupling between the field ${\phi}$ and SM fermions is suppressed by the small $\theta$ (e.g. of order $10^{-2}$) and  fermion masses, unless there is a top quark involved. As no pseudoscalar bosons beyond SM Higgs were observed at LEP, Tevatron and LHC till now, it implies that the mass of ${\phi}$ may be too heavy to be observed at present collider energies, or be within the present collider reach but very difficult to be observed. Here we would incline to the latter case. The production rate of ${\phi}$ at $e^+ e^-$, $p \bar p$ and $p \bar p$ colliders is much smaller than that of the SM Higgs boson due to the suppression of $\theta$. If the mass of ${\phi}$ is above a WIMP pair production threshold, but not heavier than the top quark, the decay rates of SM fermions would be all suppressed in ${\phi}$ decay. Even though ${\phi}$ is produced at high energy collider, the missing energy from $\phi\to \chi\bar\chi$ will be swamped by SM background. Thus, an undiscovered pseudoscalar ${\phi}$ can be consistent with the present collider experiment.

Moreover, low-energy neutral flavor-changing transitions in quark and lepton sectors will give constraints on new flavor-changing interactions. In Ref. \cite{Harnik:2012pb}, the Yukawa type flavor-changing couplings were discussed in terms of Higgs boson, and the corresponding bounds are taken here in our analysis. The ${\phi}$ mediated flavor-changing interactions can be allowed by quark and lepton sectors when $\mid \lambda_{\,i}^{} \lambda_{\,j}^{} \mid / m_\phi^2 \lesssim$ $\mid Y_{\,i} Y_{\,j}\mid / m_h^2$, where $m_\phi$, $m_h$ is $\phi$'s mass, Higgs  mass respectively, $\lambda_{\,i}^{}, \lambda_{\,j}^{}$ and $Y_{\,i}, Y_{\,j}$ are the corresponding couplings of the two vertexes connected by a ${\phi}$ boson and a Higgs boson respectively. For $\phi$'s mass above tens GeV, the case $\theta \lesssim 10^{-2}$ is allowed by the flavor-changing interactions in quark and lepton sectors, e.g. an approximate upper limit $\theta / m_\phi (GeV) \lesssim 10^{-2} / 20$ with $m_\phi$ in units of GeV.

Now let us evaluate the contribution of ${\phi}$ to top quark FCNC decays. Considering mass of ${\phi}$ to be in the range $2m_\chi<m_{\phi}<m_t$, where $m_\chi$ is the WIMP mass, according to Eq.~(\ref{SM-DM}), the leading order decay width of $t\rightarrow c \phi$ is
\begin{eqnarray}
\Gamma_{t\rightarrow c \phi}=\frac{\bar{\beta}_f}{16 \pi}m_t
 \mid \lambda_{\,tc}^{} \mid^2 [(1-\frac{m_c}{m_t})^2-\frac{m_{\phi}^2}{m_t^2}]\,, \label{tcphi}
\end{eqnarray}
where
\begin{eqnarray}
\bar{\beta}_f=\sqrt{1-\frac{2(m_{c}^2+m_{\phi}^2)}{m_t^2}+\frac{(m_{c}^2-m_{\phi}^2)^2}{m_t^4}}\,.
\end{eqnarray}
As long as the parameter $\lambda_D$ is not extremely tiny and the mass of $\phi$ is larger than the threshold of the WIMP pair, the channel of $\phi\to \chi\bar\chi$  should be overwhelming, and $\mathcal {B}_{t\rightarrow c \bar{\chi} \chi} \simeq \mathcal {B}_{t\rightarrow c \phi}$. Meanwhile, in the SM, the branching ratio of $t\rightarrow c\nu \bar{\nu}$ where neutrinos also manifest as missing energy, is of order $\sim3\times10^{-14}$ \cite{Frank:2006ku}, thus, it should be possible to identify the WIMP final state in $t\rightarrow c$ decays with missing energy.

\begin{figure}[!htbp]
\includegraphics[width=3.2in]{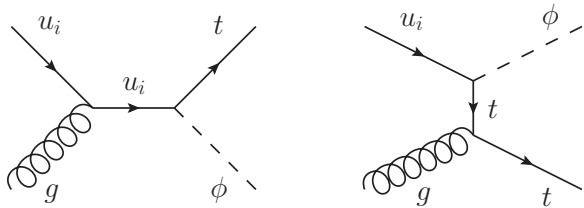} \vspace*{-1ex}
\caption{The process $u_i g \rightarrow t {\phi}$.}\label{Ui-t}
\end{figure}

The new interaction also contributes to the single top production which is a FCNC process. The fusion of a gluon and an up-type quark $u_i$ ($c$ or $u$ quark ) into a top quark and $\phi$, i.e. $u_i g \rightarrow t \phi$ at hadron collider (see Refs. \cite{Andrea:2011ws,Kamenik:2011nb,Bai:2013ooa,Greljo:2014dka} for more)  is depicted in Fig. \ref{Ui-t}. From Eq.~(\ref{SM-DM}), the flavor-changing interaction of $t-u_i-\phi$ is
\begin{eqnarray}
 \mathcal {L}_{\,{t}u_i}^{\,i} = -i \lambda_{\,{t}u_i}^{} \bar{t} \gamma_5 u_i {\phi}\,.
\end{eqnarray}
As the pseudoscalar $\phi$ is invisible, and the final products is a single top quark and a large missing energy. The recent monotop production data of the CMS Collaboration \cite{Khachatryan:2014uma} would set a constraint on the parameter for this flavor-changing coupling, and we will discuss it later.

\section{WIMP abundance and  annihilation}

Even though the relic density of DM sets constraints on the transitions between WIMPs and SM particles, in DM dense regions, the observed cosmic ray excess may shed light on the properties of WIMP annihilations. Here we are going to formulate the WIMP abundance and WIMP annihilation.

\subsection{WIMP abundance}

In the concerned model, the process of WIMP pairs annihilating into SM heavy fermions dominates.
As the width of the intermediate boson $\phi$ is smaller compared with its mass, the WIMP pair annihilation cross section can be written as
\begin{eqnarray}
\sigma_{ann}=\frac{1}{2}\sigma_{ann}^{Dirac}&=&\frac{1}{2}\frac{1}{\beta_i (2s_\chi^{}+1)^2}\frac{\beta_f}{4\pi} \lambda_{\,{f}'f}^{2} \lambda_D^2 \nonumber \\
&&\times \frac{N_c [s-(m_f-m_{f'})^2]}{(s-m_\phi^2)^2+m_\phi^2\Gamma_\phi^2} \,. \label{dm-ann}
\end{eqnarray}
The factor $\frac{1}{2}$ in the expression is for the Dirac fermionic WIMPs whereas it is 1 for the Majorana fermionic WIMPs.
$s$ is the square of the total invariant mass, and $s_\chi$ is the WIMP spin projection. $\Gamma_\phi$ is the total width of $\phi$, and $N_c$ is the color factor of the final state fermions. The kinematic factors $\beta_i$, $\beta_f$ are
\begin{eqnarray}
\beta_i&=&\sqrt{1-\frac{4 m_\chi^2}{s}}, \, \nonumber \\ \beta_f&=&\sqrt{1-\frac{2(m_f^2+m_{f'}^2)}{s}+\frac{(m_f^2-m_{f'}^2)^2}{s^2}}.
\end{eqnarray}

We get the WIMP annihilation cross section by Eq. (\ref{dm-ann}). The DM relic density $\Omega_D$ depends on the
thermal evolution after the Big-Bang. The values of the relic density and
the freeze-out temperature can be approximately written as \cite{Kolb:1990vq,Griest:1990kh}
\begin{eqnarray}
\Omega_{D} h^2\simeq \frac{1.07\times10^9 x_f}{\sqrt{g_\ast}m_{\rm Pl}GeV\langle\sigma_{ann}v_{rel}\rangle} \,,
\end{eqnarray}
\begin{eqnarray}
x_f\simeq \ln \frac{0.038~ g m_{\rm Pl}m_{\chi}\langle\sigma_{ann}v_{rel}\rangle}{\sqrt{g_\ast x_f}} \,. \label{xf}
\end{eqnarray}
Here $h$ is the Hubble constant in units of 100 km/($s \cdot$ Mpc). $x_f^{}$ is equal to $m_\chi^{}/T_f^{}$ with $T_f^{}$ being the freeze-out temperature, and $g^{}_{*}$ is the number of the relativistic degrees of
freedom for masses being less than $T_f^{}$. $m_{\rm Pl}^{}=1.22 \times 10^{19}$ GeV is the Planck mass, and the number $g$ is the degrees of freedom of DM. $\langle \sigma_{ann} v_{rel} \rangle$ is the thermal-averaged annihilation cross section of WIMPs $\to$ SM particles, and $v_{rel}$ is the relative speed of the annihilating WIMP pair. The thermal average of the annihilation cross section is \cite{Gondolo:1990dk}
\begin{eqnarray}
\langle \sigma_{ann} v_{rel} \rangle&=&\frac{1}{8 m_{\chi}^4 T K_2^2(\frac{m_{\chi}}{T})}  \int_{4 m_{\chi}^2}^\infty \rm{d} s~   \nonumber \\
&&\times \sigma_{ann} \sqrt{s} (s - 4 m_{\chi}^2) K_1(\frac{\sqrt{s}}{T})\,,
\end{eqnarray}
where $K_i(x)$ is the $i-$th order modified Bessel function.

The value of $x_f$ can be obtained by solving the Eq. (\ref{xf}) iteratively. $g_\ast$ varies with the freeze-out temperature $T_f$, and we will adopt the MicrOMEGAs 3.1 data \cite{Belanger:2013oya} about the Gondolo-Gelmini effective degrees of freedom at $T_{QCD}$ = 150 MeV in numerical calculations.

\subsection{Present WIMP annihilation rate}

As the today's environmental temperature is negligible compared with the WIMP mass, the annihilation rate of WIMPs is derived in the $T=0$ limit. From Eq. (\ref{dm-ann}), the present WIMP annihilation rate is
\begin{eqnarray}
\langle \sigma_{ann} v_r \rangle=\langle \frac{1}{2}  \sigma_{ann}^{Dirac} \, v_r \rangle_{T=0}&=&\frac{1}{2} \frac{1}{(2s_\chi^{}+1)^2}\frac{\beta_f}{2\pi} \lambda_{\,{f}'f}^{2} \lambda_D^2 \nonumber \\
&&\times  \frac{ N_c[4 m_{\chi}^2-(m_f-m_{f'})^2]}{(4 m_{\chi}^2-m_\phi^2)^2+m_\phi^2\Gamma_\phi^2} \,, \label{Today-ann}
\end{eqnarray}
with the same notations as given in Eq. (\ref{dm-ann}). $v_r$ is the relative speed of the WIMP pair, and here $s=4 m_{\chi}^2$ has been taken due to the small velocity of WIMPs \cite{Smith:2006ym}.

In the range $2m_\chi<m_{\phi}<m_t$, if the mass $m_{\chi}$ is heaver than b quark, the dominant products of WIMP pairs annihilating into SM particles are $b \bar b$ quarks, and this coincides with the observed GeV gamma ray excess and the corresponding interpretation.

The present WIMP annihilation rate $\langle \sigma_{ann} v_r \rangle$ is not equal to the value of $\langle \sigma_{ann} v_{rel} \rangle$ at freeze-out temperature (see e.g. \cite{Kozaczuk:2015bea}), especially when the value of $2m_\chi$ is around $ m_{\phi}$. Generally, when the WIMP pair mass $2m_\chi$ is below the mass of $\phi$, we have $\langle \sigma_{ann} v_r \rangle <$ $\langle \sigma_{ann} v_{rel} \rangle$, while $\langle \sigma_{ann} v_r \rangle >$ $\langle \sigma_{ann} v_{rel} \rangle$ for $2m_\chi >$ $ m_{\phi}$. In the next section, we will numerical analysis the WIMP pair annihilation in detail.

\section{Numerical analysis of $t - c$ transitions}

Here, we first briefly analyze the single top production, then focus on the top quark decay.

\subsection{Single top production}

The data on single top production at hadron collider set certain constraints on $c\rightarrow t$ transitions. The CMS Collaboration's results \cite{Khachatryan:2014uma} set an upper limit on the flavor-changing couplings $a_{FC}^0$ in the process $u_i g \rightarrow t \phi_0$, with $\phi_0$ (we deliberately add a subscript $0$ to distinguished it from $\phi$) being a spin-0 invisible boson. If only considering $\phi_0$ contributing to the u-t transition with the coupling $a_{FC}^0= 0.1$, the observed data well fit the SM expectations with no excess signal being observed at 95\% confidence level, and a new particle $\phi_0$ with mass below 330 GeV is excluded.

This constraint can be relaxed for a pseudoscalar ${\phi}$ which mediates flavor-changing interaction. If $\theta \sim 10^{-2}$ is taken, the c-t coupling $\lambda_{\,t c}^{}$ is of order $10^{-3}$. The u-t coupling $\lambda_{\,t u}^{}$ should be even smaller and is about $3.6 \times 10^{-5}$ with a small current mass of $m_u\sim$ 2.3 MeV. Thus,  the results obtained by the CMS Collaboration (see Fig. 3 of Ref. \cite{Khachatryan:2014uma}) imply that the contribution of ${\phi}$  to the single top production is very small compared with that of SM, and the pseudoscalar ${\phi}$-relevant signals are swamped by the SM background.

In the energy range of concern, the monotop production at LHC may not be employed for determining the new flavor-changing interactions, and namely any solid
information about WIMPs cannot be extracted from the measurements on the corresponding missing energy.

\subsection{$t\to c$ decay}

Let us now turn to the $t\to c$ decay which was discussed in the introduction as an optimistic channel for detecting WIMPs, and here consider the mass of WIMPs in the range 5-60 GeV. The $b \bar{b}$ pairs are the dominant products in WIMPs annihilation into SM, and very few $c \bar{c}$, $\tau \bar{\tau}$, $b \bar{s}$, etc. may take a small fraction. The relic abundance of cold DM observed today is employed to restrict the parameter spaces.

\subsubsection{Constraints set by the WIMP annihilations}

The relic density of cold DM today is $\Omega_{c} h^2=0.1197 \pm 0.0022$  \cite{Planck:2015xua}, and this value can be taken to restrict the parameter spaces. We set $m_t=173.21$ GeV, $m_b(\overline{MS}) = 4.18 $ GeV, $m_c = 1.275$ GeV, $m_s = 0.095$ GeV, $m_{\tau} = 1.7768$ GeV according to the Particle Data Group values \cite{Agashe:2014kda}.

\begin{table}[!htbp]
\caption{The numerical results of $\xi_c$, for $\lambda_D =$0.5, 1.0, $m_\chi =$ 5, 20, 35, 50, 60 GeV.} \label{Table1}
\begin{tabular}{r c c c c c}
  \hline
  \hline
  $m_\chi$(GeV) & 5 & 20 & 35 & 50 & 60 \\
  \hline
  $\xi_c$($\lambda_D=$0.5) & 0.9971 & 0.9971 & 0.9972 & 0.9972 & 0.9973 \\
   $\theta_c$($\xi_c$)$\times 10^{-2}$ & 0.107 & 0.330 & 0.562 & 0.795 & 0.950 \\
   \hline
  $\xi_c$($\lambda_D=$1.0)& 0.9940 & 0.9939 & 0.9940 & 0.9941 & 0.9941 \\
  $\theta_c$($\xi_c$)$\times 10^{-2}$ & 0.120 & 0.374 & 0.638 & 0.903 & 1.08 \\
  \hline
\end{tabular}
\end{table}

Let us define $2m_\chi / m_{\phi} = \xi$, with $\xi < 1$ here. When the mass $m_{\phi}$ is slightly above $2m_\chi$, the WIMP pair annihilation rate will be enhanced. Using WIMP relic density, we can derive the thermal averaged annihilation cross section $\langle \sigma_{ann} v_{rel} \rangle$ of concern is about $(1.63-1.76)\times 10^{-9}$ GeV$^{-2}$ ($\sim (1.91-2.06)\times 10^{-26} $ cm$^3/$s). There is a critical value $\xi_c$, when $\xi \gtrsim \xi_c$, the annihilation cross section $\langle \sigma_{ann} v_r \rangle_{b\bar{b}}$ (the dominant products $b\bar{b}$ in SM) at $T=0$ will be not smaller than $\langle \sigma_{ann} v_{rel} \rangle$ at $T=T_f$. The numerical results of $\xi_c$ are shown in Table \ref{Table1}, for $\lambda_D =$0.5, 1.0, $m_\chi =$ 5, 20, 35, 50, 60 GeV. The new dwarf galaxy $b \bar{b}$ channel annihilation limits \cite{Ackermann:2015zua} show that, $\langle \sigma_{ann} v_r \rangle_{b\bar{b}} \lesssim \langle \sigma_{ann} v_{rel} \rangle$, for WIMPs with masses below a hundred GeV.  Thus, $\xi_c$ can be taken as an upper bound of $\xi$, i.e. $\xi \lesssim \xi_c$.

\begin{table}[!htbp]
\caption{The numerical results of $\xi_l$, for $\theta_u= \theta_{m_\chi}, 1.0 \times 10^{-2}$, $\lambda_D =$0.5, 1.0, $m_\chi =$ 5, 20, 35, 50, 60 GeV. The `null' in the table means no parameter space.} \label{Table2}
\begin{tabular}{r c c c c c}
  \hline
  \hline
  $m_\chi$(GeV) \quad & 5 & 20 & 35 & 50 & 60 \\
  \hline
  \multirow{2}{*}{$\theta_u= \theta_{m_\chi}$}  $\xi_l$($\lambda_D=$0.5)  & 0.9164 & 0.9126 & 0.9135 & 0.9144 & 0.9181 \\
    $\xi_l$($\lambda_D=$1.0)  & 0.9147 & 0.9095 & 0.9103 & 0.9111 & 0.9117 \\
  \hline
  \multirow{2}{*}{$\theta_u= 0.010$} $\xi_l$($\lambda_D=$0.5) & 0.8838 & 0.9436 & 0.9694 & 0.9862 & 0.9948 \\
     $\xi_l$($\lambda_D=$1.0)  & 0.8775 & 0.9439 & 0.9716 & 0.9892 & null \\
  \hline
\end{tabular}
\end{table}

To meet the DM relic density observed, there is a $\xi_c$-dependent parameter $\theta_c$, and $\theta_c$ is determined by the case of $\xi \sim \xi_c$, $\theta \sim \theta_c$. The results of $\theta_c (\xi_c)$ are also given in Table \ref{Table1}, for the set WIMP masses and $\lambda_D$. It can be seen that $\theta_c$ varies within a range of $10^{-3} - 10^{-2}$, and $\theta_c$ is taken as the lower bound of $\theta$, i.e. $\theta \gtrsim \theta_c$. Meanwhile, $\theta$ needs to be as small as possible associated with SM. An upper limit $\theta \lesssim \theta_u$ with $\theta_u$ of order $10^{-2}$ is of concern, and correspondingly, there is a $\theta_u$-dependent lower limit $\xi_l$ for $\xi$, $\xi \gtrsim \xi_l$. Considering the low-energy flavor-changing constraints in quark and lepton sectors discussed above, we have
\begin{eqnarray}
\frac{\xi \, \theta}{2m_\chi (GeV)}  \lesssim \frac{10^{-2}}{ 20}.
\end{eqnarray}
Noting a $m_\chi$-dependent parameter $\theta_{m_\chi}$, with
\begin{eqnarray}
\theta_{m_\chi} = 10^{-2} \times \frac{m_\chi  (GeV)}{10},
\end{eqnarray}
thus $\theta_u \sim \theta_{m_\chi}$ can be taken as an upper limit. The numerical results of $\xi_l$ are given in Table \ref{Table2} for $\theta_u= \theta_{m_\chi}$, along with $\theta_u=1.0 \times 10^{-2}$ as comparison.

\begin{figure}[!htbp]
\includegraphics[width=3.6in]{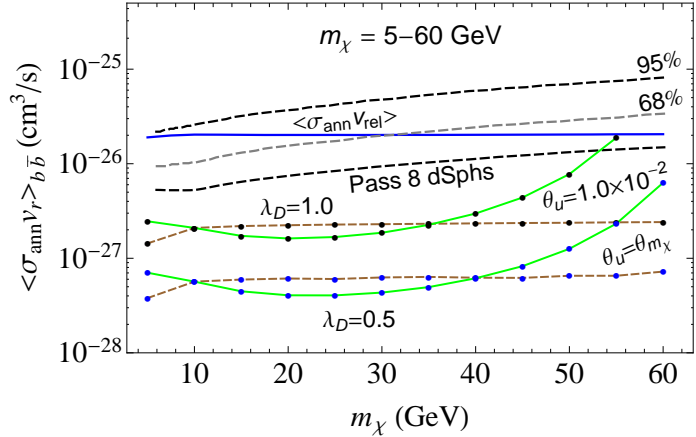} \vspace*{-1ex}
\caption{The WIMP annihilation cross section $\langle \sigma_{ann} v_r \rangle_{b\bar{b}}$ as a function of $m_\chi$, for $m_\chi$ varying in the range $5 - 60$ GeV. The dashed-dotted, solid-dotted curves are for the case of $\theta =$ $\theta_u \sim$ $\theta_{m_\chi}$, $1.0\times 10^{-2}$ respectively, with the upper one, lower one in each type two curves corresponding to $\lambda_D=$1.0, 0.5 respectively. The solid curve is the WIMP annihilation cross section $\langle \sigma_{ann} v_{rel} \rangle$. The dashed curves are the Pass 8 dSphs upper limits of $b\bar{b}$ from the Fermi-LAT Collaboration \cite{Ackermann:2015zua}, for the combined limit, $68 \%$ containment, $95 \%$ containment respectively from bottom to top.}\label{WIMP-annv}
\end{figure}

When the parameters $\xi$, $\theta$ vary in the range $\xi_c \gtrsim \xi \gtrsim \xi_l$, $\theta_c \lesssim \theta \lesssim \theta_u$ respectively to satisfy the observed DM relic density, the WIMP annihilation cross section $\langle \sigma_{ann} v_r \rangle_{b\bar{b}}$ is also determined. The result of $\langle \sigma_{ann} v_r \rangle_{b\bar{b}}$ is depicted in Fig. \ref{WIMP-annv}, for WIMP masses varying from 5 to 60 GeV, and $\theta =$ $\theta_u \sim$ $\theta_{m_\chi}$, $1.0\times 10^{-2}$, $\lambda_D=$1.0, 0.5 respectively. The coupling $\lambda_D$ dominates the width $\Gamma_\phi$, which is important near the resonance in WIMP annihilations. In the case $\theta \sim $ $\theta_c$, $\langle \sigma_{ann} v_r \rangle_{b\bar{b}}$ approximately equals to $\langle \sigma_{ann} v_{rel} \rangle$. It can be seen that, considering the Pass 8 dSphs (dwarf spheroidal satellite galaxies) upper limits of $b\bar{b}$ from the Fermi-LAT Collaboration \cite{Ackermann:2015zua}, there are still parameter spaces allowed.

As pointed by the analysis of Ref. \cite{Hooper:2015ula}, the new dwarf galaxy limits \cite{Ackermann:2015zua} can be compatible with the galaxy center excess. In this case,
when the value of $\theta$ is about $\theta_c$, the corresponding $\langle \sigma_{ann} v_r \rangle_{b\bar{b}}$ is $\sim 10^{-26}$ cm$^3/$s, which can fit the GeV gamma ray excess.

\subsubsection{The decay rate}

Now we turn to the WIMP production in $t\to c$ decay with the allowed parameter spaces. The ranges of parameters $\theta$, $\xi$ determine the decay width $\Gamma_{t\rightarrow c \bar{\chi} \chi}$. According to Eq.~(\ref{tcphi}) and neglecting  ${m_c}/{m_t}$, the decay width is rewritten as
\begin{eqnarray}
\Gamma_{t\rightarrow c \bar{\chi} \chi} \simeq \frac{1}{16 \pi}m_t
 \mid \lambda_{\,tc}^{} \mid^2 (1-\frac{4 m_{\chi}^2}{\xi^2 \, m_t^2})^2\,.
\end{eqnarray}

\begin{figure}[!htbp]
\includegraphics[width=3.6in]{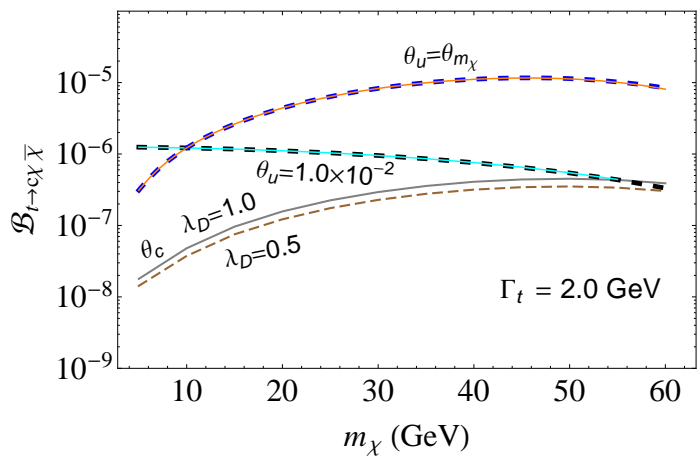} \vspace*{-1ex}
\caption{The branching ratio $\mathcal {B}_{t\rightarrow c \bar{\chi} \chi}$ as a function of $m_\chi$, for $m_\chi$ varying from 5 to 60 GeV. $\Gamma_t = $ 2.0 GeV is taken. The dashed, solid curves   correspond to $\lambda_D=$ 0.5, 1.0 respectively, for both cases we set $\theta =$ $\theta_c $. The dashed and solid overlapping curves are for the case of $\theta =$ $\theta_u$ (for the dashed, solid part, $\lambda_D=$ 0.5, 1.0 respectively), with the upward bending, the downward bending curves corresponding to $\theta_u =$ $\theta_{m_\chi}$, $1.0\times 10^{-2}$ respectively.}\label{brtc}
\end{figure}

The total top quark decay width is $2.0 \pm 0.5$ GeV \cite{Agashe:2014kda}. Substituting the ranges of $\theta$, $\xi$, we can obtain the branching ratio $\mathcal {B}_{t\rightarrow c \bar{\chi} \chi}$, as
shown in Fig. \ref{brtc}, where $m_\chi$ varies within the range 5 GeV $\leq m_\chi \leq$ 60 GeV, and $\Gamma_t = $ 2.0 GeV is taken. The case of $\theta = \theta_c$ is adopted as the lower limit, with $\lambda_D=$ 0.5, 1.0. The case of $\theta = \theta_u =$ $\theta_{m_\chi}$, $1.0\times 10^{-2}$ is taken as the upper limit, and for each $\theta_u$, the curves with $\lambda_D=$ 0.5, 1.0 are overlapped. There is no parameter space for $m_\chi \gtrsim $ 55 GeV with $\theta_u = 1.0\times 10^{-2}$ and $\lambda_D=$ 1.0. When $\theta$ varies between $\theta_c $ and $ \theta_u$, the the branching ratio $\mathcal {B}_{t\rightarrow c \bar{\chi} \chi}$ is of order $10^{-8} - 10^{-5}$. There is about 25\% uncertainty for the branching ratio, which is mainly coming from the value of top quark decay width.

The $t\rightarrow c \bar{\chi} \chi$ decay can be searched at LHC, e.g. by the process $p p \rightarrow t \bar t$ $\rightarrow c \phi + \bar{b} W^- $ with missing energy, and such process was discussed in Ref. \cite{Li:2011ja}. There are about $10^6$ order $t \bar t$ pairs production \cite{Khachatryan:2014loa,Aad:2015pga} at 8 TeV center-of-mass energy at LHC Run I, and maybe there are a few events produced. For LHC run at 14 TeV, the top quark pair production cross-section is about 953.6 pb \cite{Czakon:2013goa}. As indicated in Ref. \cite{Li:2011ja}, for the branching ratio $t\rightarrow c \bar{\chi} \chi$ of order $10^{-5}$, the decay mode seems challenging to be observed at LHC next run.

The WIMP signature may appear in $t\rightarrow c$ decay with missing energy at high energy collider, or the corresponding upper limit of the $\theta$ is set by experiments. The observation of cosmic ray excess from DM dense region indirectly gives the property of WIMP annihilations, or raises the lower limit of $\theta$ (meanwhile, it may also give constraints on the upper limit of $\theta$ as well). Thus, the pseudoscalar $\phi$ mediated WIMPs will be examined by the joint effort of collider search and cosmic ray observation.

\section{Conclusion and discussion}

The scenario of a pseudoscalar $\phi$ mediating FCNC Yukawa type interaction which results in a transition of $t\to c$ is discussed in this work. In this model, the WIMP-nucleus scattering is highly suppressed, so that one cannot expect to ``see" them via direct detection, and the small $\theta$ value and fermion masses (except top quark) let the new interactions be tolerant by the present observation at LHC. In the case of $2m_\chi<m_{\phi}<m_t$, the SM channels in $\phi$ decay are suppressed, and the $\bar{\chi} \chi$ channel is dominant. The missing energy in the process of the single top production and $t\to c$ decay is employed to search for WIMPs. For the single top production with missing energy at hadron collider, we notice that the contribution of $\phi$ is swamped in the SM background when $\theta \sim 10^{-2}$.

We consider the mass of WIMPs in the range 5-60 GeV, and the dominant products of WIMPs annihilating into SM particles are $b \bar{b}$ pairs, and this scenario is favored by the discovery of a gamma ray excess from the galaxy center. The WIMP relic abundance is employed to restrict the parameter spaces. We find that considering the constraints of Pass 8 dSphs upper limits of $b\bar{b}$ \cite{Ackermann:2015zua}, there is parameter space left for $\langle \sigma_{ann} v_r \rangle_{b\bar{b}} \lesssim$ $\langle \sigma_{ann} v_{rel} \rangle$, and the GeV gamma ray excess can be fitted within the parameter space. Thus, our numerical analysis about WIMP pair annihilations is still effective even if the $b \bar{b}$ mode no longer fits the Galactic Center gamma-ray excess. With the parameter spaces allowed,  we derive the branching ratio $\mathcal {B}_{t\rightarrow c \bar{\chi} \chi}$ which is of order $10^{-8} - 10^{-5}$.

The $t\rightarrow c \bar{\chi} \chi$ decay can be explored at LHC and the proposed ILC, while its observation seems challenging at LHC next run. The much clean environment for $t \bar t$ production at ILC \cite{Baer:2013cma} and FCC-ee (TLEP) \cite{Gomez-Ceballos:2013zzn} will offer incredible opportunity to measure the $t\rightarrow c \bar{\chi} \chi$ decay. For $e^+ e^-$ collider at 500 GeV, the $t \bar t$ production cross-section is about 0.572 pb \cite{Amjad:2013tlv}, and thus the branching ratio $t\rightarrow c \bar{\chi} \chi$ of order $10^{-5}$ can be searched with an integrated luminosity of 1000 fb$^{-1}$. Thus, the excess favored WIMPs from the  galaxy center could be explored at future $e^+ e^-$ collider. Even though one cannot expect to gain definite conclusion, measurement on $t\rightarrow c $ transitions with missing energy at collider can help to reduce the upper limit of $\theta$. On other aspect, whereas the indirect observation of WIMP annihilations will raise the lower limit of $\theta$.

Moreover, when the WIMP pair mass is above the top quark mass, the $t \bar{c}$ ($\bar{t} c$) mode is opened in WIMP pair annihilations, and another alternative interpretation of the gamma ray excess is WIMP pair annihilating into top-charm quark mode \cite{Rajaraman:2015xka}. In this case, there is still parameter space for the case $m_t + m_c < 2 m_\chi $  $< 2 m_t$ to satisfy the constraints, while the WIMPs searching in top quark decay is impossible.

Though the pseudoscalar $\phi$ mediated WIMPs could evade our direct detection for the highly suppressed WIMP-nucleus scattering, with the joint efforts of collider search, cosmic ray observation, and other means, the scenario where pseudoscalar $\phi$ mediates interaction between WIMPs and SM particles will be tested.

\acknowledgments \vspace*{-3ex}  Thank Xue-Qian Li for helpful discussion. This work was supported by the Research Fund for the Doctoral Program of Southwest University of Science and
Technology under the contract No. 15zx7102.

\end{document}